\documentclass[final]{elsart}

\usepackage{color}
\definecolor{grey}{gray}{.75}
\definecolor{red}{rgb}{1,0,0}

\usepackage{epsfig}
\usepackage{slashbox}
\usepackage{gensymb}

\usepackage{lineno}
\usepackage{amssymb}
\usepackage{amsmath}

\begin{document}

\begin{frontmatter}



\title{Implementation of mean-timing and subsequent logic functions on an FPGA}


\author{J. Bieling,}
\author{G. Ahluwalia,}
\author{J. Barth,}
\author{F. Klein,}
\author{J. Pretz\corauthref{cor1}}
\corauth[cor1]{corresponding author}
\ead{pretz@physik.uni-bonn.de}
\address{Physikalisches Institut, Universit\"at Bonn, 53115 Bonn, Germany}
\author{H. Fischer,}
\author{F. Herrmann,}
\author{C. Schill,}
\author{S. Schopferer}
\address{Physikalisches Institut, Universit\"at Freiburg, 79104 Freiburg, Germany}

\begin{abstract}
This article describes the implementation of a mean-timer 
and  coincidence logic on a Virtex-5 FPGA for trigger purposes
in a particle physics experiment.
The novel feature is that the mean-timing and the coincidence logic are not synchronized 
with a clock which allows for a higher resolution of approximately 400 ps, not
limited
by a clock frequency.
\end{abstract}

\end{frontmatter}

% main text

%\textin{}
%\textout{}
%-------------------------------------------------------------------------------
\section{Introduction}
For scattering experiments in particle physics one often needs fast trigger
decisions. 
% of the order of 500ns, to trigger the data acquisition system.
The trigger signal should have a good time resolution with respect to 
the occurrence of the scattering event, of the order of 1~ns.
Scintillation detectors allow for fast trigger decisions
with an appropriate time resolution.
However, in large scintillation detectors the time resolution is deteriorated by a
varying light propagation time depending on the impact point of a particle
in the detector.
Thus, such long scintillator strips are read out on both sides 
by photo multiplier tubes (PMTs), and 
the method of mean-timing
is employed by generating in real time a signal whose time corresponds to the
average time of the individual PMT pulses and is thus independent of the
impact point.
Various methods have been developed to meet this requirement
~\cite{tdl,current_source,Sandberg:1985yc}.

In this article an implementation of a mean-timer logic
and a subsequent coincidence logic is presented.
The goal was to use an already existing multi-purpose FPGA
board equipped with a Xilinx Virtex-5 allowing for a maximum clock frequency
of 500 MHz. The corresponding granularity of 2 ns was not enough for our
application, so we decided to use an unclocked logic which 
does not underlie the discreteness 
of the clock cycles.
Up to now such a mean-timer was only implemented on a CPLD~\cite{mt_cpld}.

The paper is organized as follows.
Section~\ref{prb} describes the requirements of the system
and the FPGA board used.
The main Section~\ref{impl} presents the implementation of the
mean-timer and coincidence logic on the FPGA board.
Section~\ref{perf} discusses the implementation and performance 
of the system at the COMPASS experiment at CERN~\cite{compass}.

\section{Requirements of the system and the FPGA board}\label{prb}

\subsection{The COMPASS Trigger System}
\begin{figure}
\includegraphics[width=\textwidth]{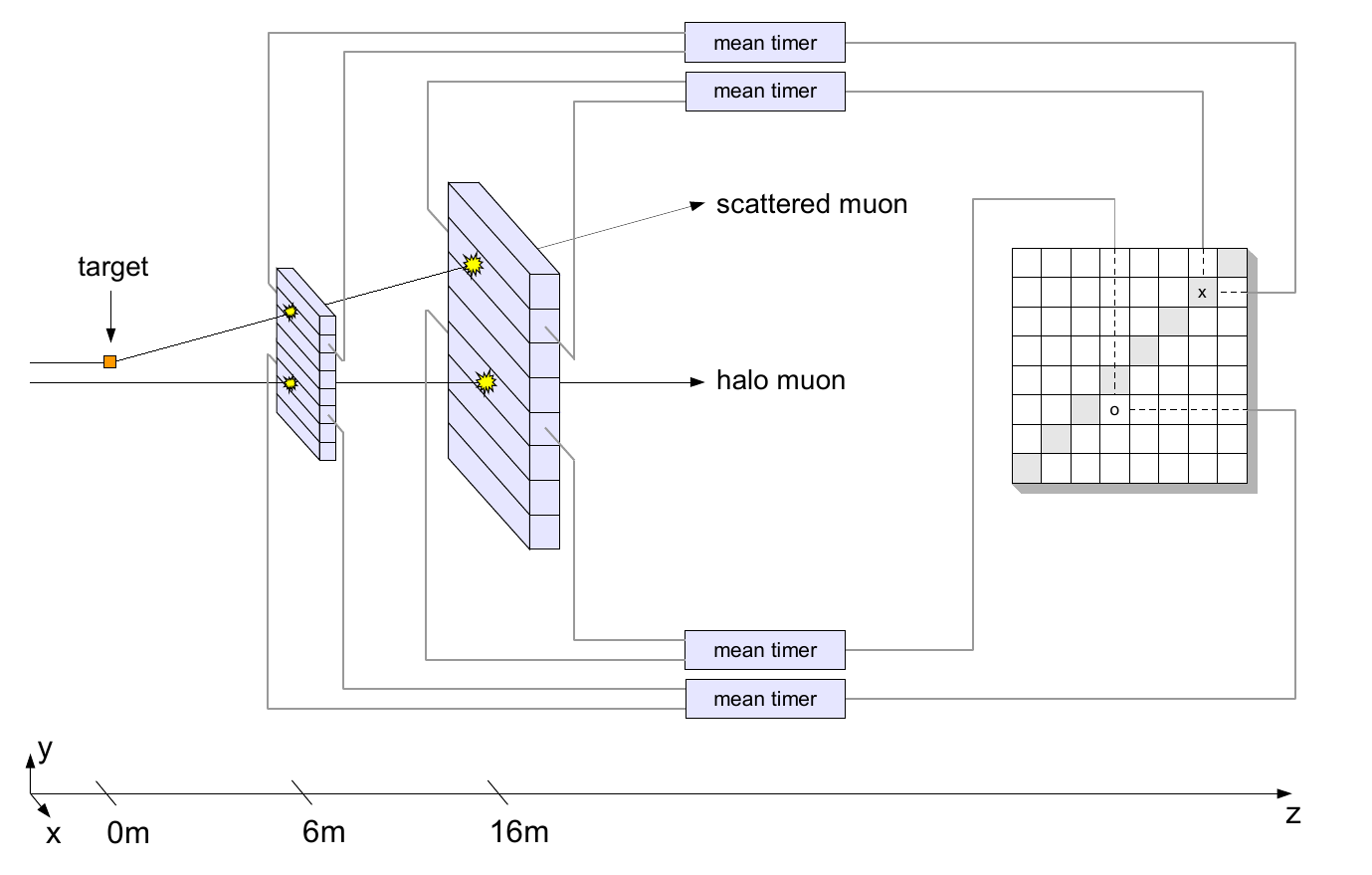}
\caption{Trigger principle. Particles are detected in two scintillator
  hodoscopes placed at different positions along the beam axis.
The 250~cm long scintillator strips are read out on both sides with photo
multiplier tubes (PMTs). After conversion to a logical signal, using discriminators, the
mean-timing is done. The output signals of the mean-timers are sent
to a coincidence matrix unit. 
Selection of the coincidences (grey squares) allows to trigger on particles
originating from the target region and to suppress background events
originating from beam halo particles.\label{trigger}}
\end{figure}

Fig.~\ref{trigger} explains the main features of the trigger
system at the COMPASS experiment~\cite{compass_trigger}.
Muons produce light in two scintillator hodoscopes placed along the beam direction.
The scintillator strips are read out on both sides by PMTs. 
After conversion to logic signals, these are passed on to the mean-timer.
For offline analysis, the single PMT signals are digitized by 
time-to-digital-converters (TDCs).

In a subsequent coincidence logic, the mean-timer outputs of 
the two hodoscopes are passed on to a coincidence matrix
which allows to select coincidences between the strips of the two hodoscopes. By selecting
only certain channel combinations of H1 and H2, as indicated in Fig.~\ref{trigger}, it is possible to discriminate 
scattered particles from beam halo particles. 
Both hodoscopes have 32 strips, so in total 64 mean-timers and a
32$\times$32 coincidence matrix are needed. Since the length of each scintillator strip is 250~cm, 
which corresponds to a light propagation time of approximately 20~ns, the dynamic range of each mean-timer 
is required to be at least $\pm 20$~ns.

\subsection{The FPGA Board}\label{board}
\begin{figure}[ht]
	\centering
	\includegraphics[width=0.8\textwidth]{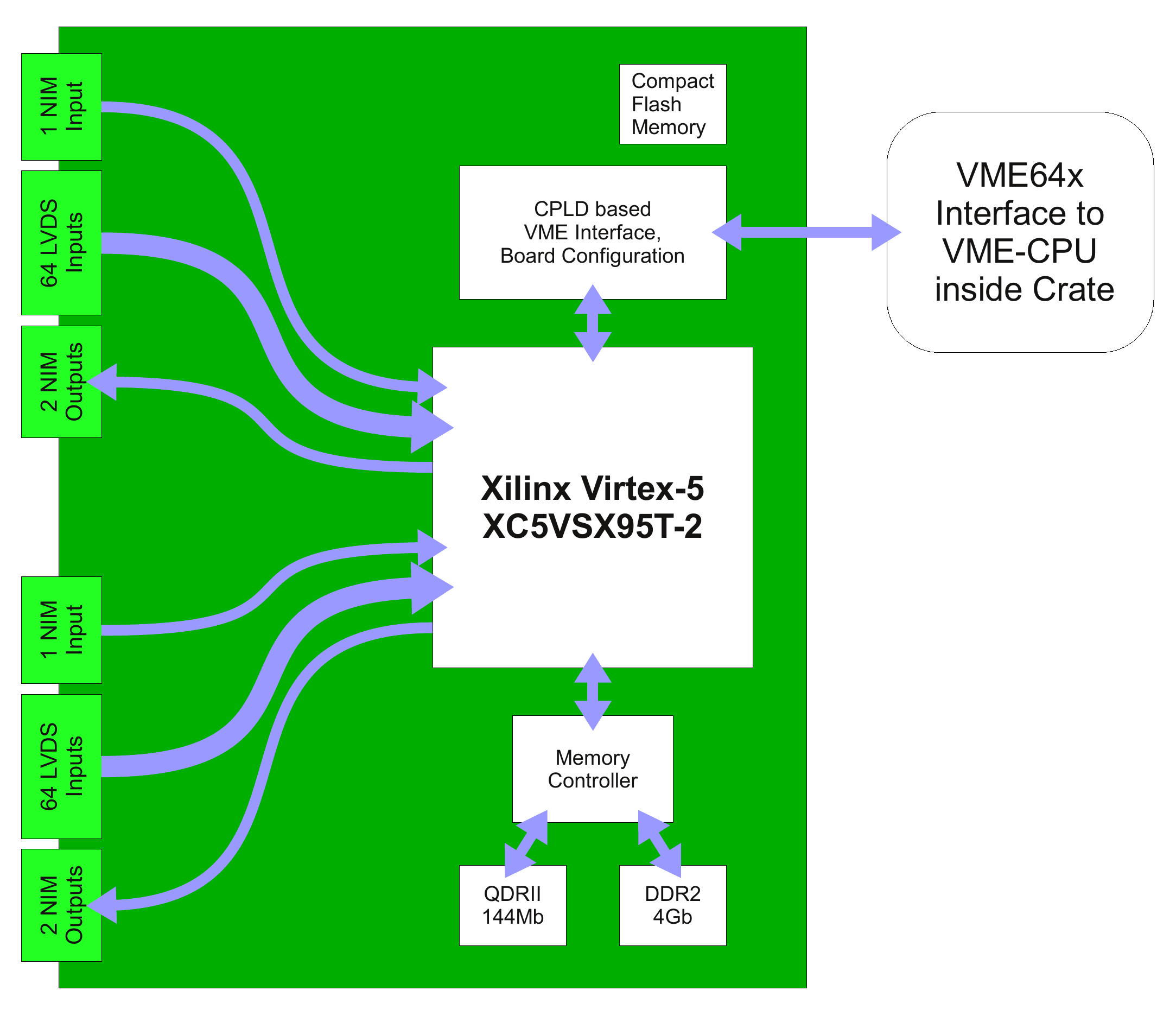}
	\caption{Simplified layout of the GANDALF-Board.}
	\label{fig:gandalf}
\end{figure}

%\textout{It turned out that the realization of the project relied heavily on the layout
%of the FPGA module. In the following ???....} 
For the implementation of the mean-timer and the coincidence logic,
we use a custom made FPGA board developed at the University of Freiburg \cite{gandalf}.
This so called GANDALF board can be equipped with two input cards providing 2 $\times$ 64 LVDS\footnote{Low Voltage Differential Signal} 
inputs and in addition 2 NIM\footnote{Nuclear Instrumentation Module} inputs and 4 NIM outputs (see Fig.~\ref{fig:gandalf}). 
The board is constructed as a 6U-VME64x/VXS module and can be controlled via the VME-Bus.
It is based on a Xilinx Virtex-5 (XC5VSX95T-2) FPGA~\cite{xilinx}. 
This FPGA consists of an array of $160 \times 46$ configurable logic blocks (CLB). 
Each of these CLBs contains 8 programmable logic elements and various other elements like a carry logic
and flip-flops. The programmable logic elements are called look-up tables 
(LUTs) and have 6 inputs and 2 outputs, so that any 6-to-2 logic can be implemented. 
To allow communication between different LUTs and flip-flops, every CLB has a switch matrix to which 
all elements of a CLB are connected. If two elements reside in different CLBs their switch matrices use 
so called global interconnects.
The FPGA provides 640 I/O connections, from which 256 are used for the 128
LVDS inputs, and 6 for the NIM inputs and outputs. 
To each of these connections belongs a dedicated delay element which allows to delay the signal in 64 steps of 75~ps. 
Thus the maximum delay is about 5~ns.
A 36~kbit block-Ram, a dynamic clock manager (DCM) and some CLBs optimized for digital signal processing algorithms 
complete the structure of the FPGA. 

Due to variations in the production process, each FPGA has its own timing behavior. To be able to design 
clocked circuits, 
however, some maximum propagation timings must not be exceeded. Therefore, Xilinx characterizes every single 
FPGA by its timing behavior by so
called speed grades ranging from 1 to 3. A speed grade of 1 corresponds to the
slowest version~\cite{speed_grade}. In the simulation software the developer has to select the correct 
speed grade. It is indicated by the last digit in the device number, in our case an
FPGA with speed grade 2 is used.

\section{Implementation of Mean-Timer and Coincidence Logic}\label{impl}
\subsection{Implementation of Mean-Timers}\label{mt}

\begin{figure}
\includegraphics[width=\textwidth]{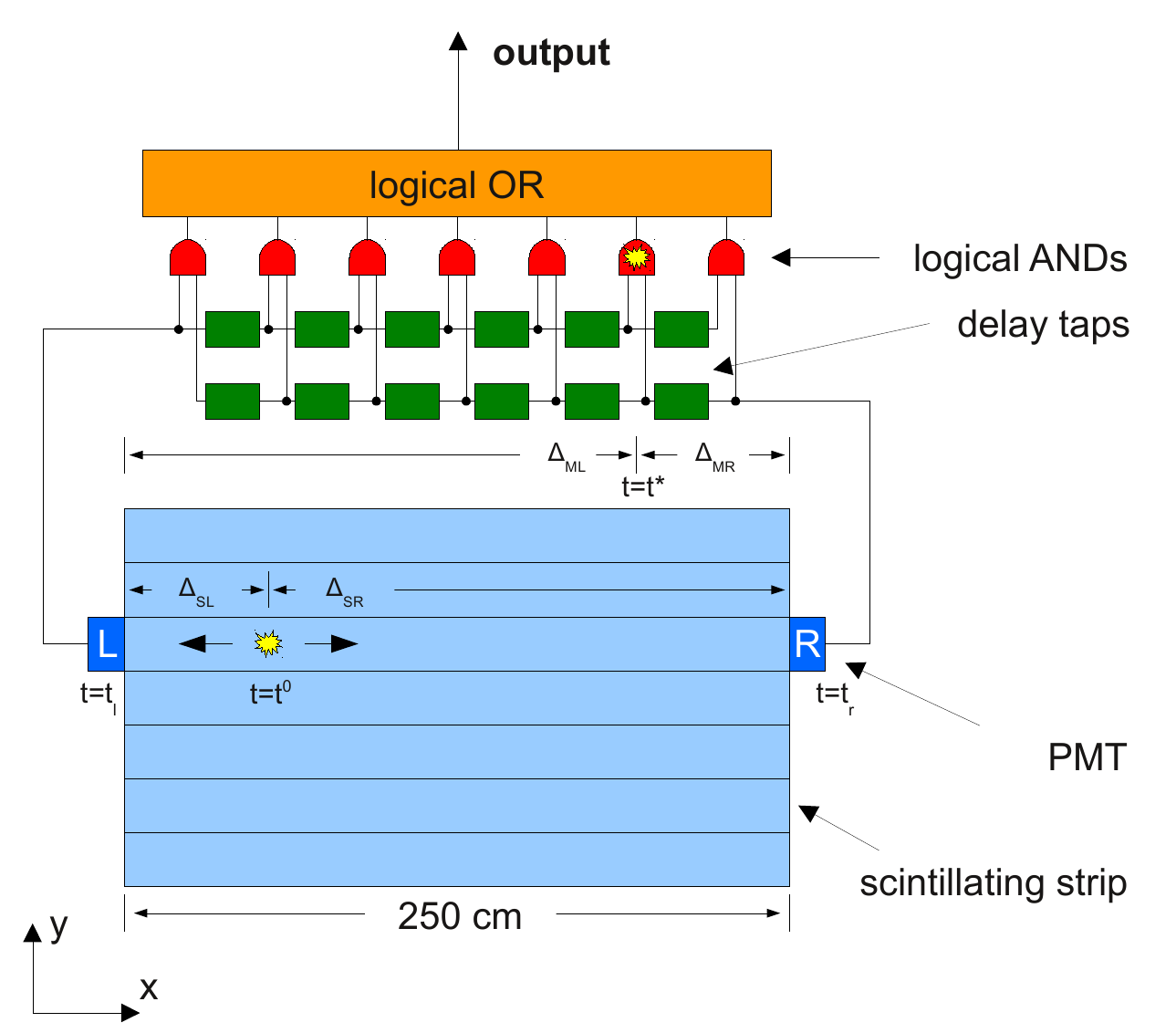}
\caption{Principle of the mean-timer using a tapped delay line. \label{meantimer}}
\end{figure}

The mean-timers are constructed using the tapped delay line (TDL) method~\cite{tdl} as shown in Fig.~\ref{meantimer}
and App.~\ref{app_tdl}.
The two input signals of each mean-timer propagate in opposite directions in the two delay lines and trigger 
the ANDs between these lines while passing each other. The
OR-ed output of all ANDs provides 
the desired mean-timed signal. 
Implementing an unclocked mean-timer on a FPGA, one 
has to take care that 
\begin{enumerate}
\item all delays on the tapped delay line are of equal amount
(up to a few ps) and
\item the delays from the ANDs to the logical OR are identical.
\end{enumerate}

A chain of LUTs is used to build the tapped delay line.
It turns out that the delays between the 8 LUTs inside the same CLB are much
smaller than the delays 
between two LUTs located in two different CLBs, 
because the second case requires global interconnects. 
Since more than 8 delay elements are needed to reach the desired dynamic
range, the tapped delay lines have to be constructed using those
interconnects. To determine the path length of a particular interconnect
between two LUTs, we used LOC-Constraints (location constraints) to place the LUTs
at certain positions and added a signal path from one of the two outputs from the first LUT to one of the
6 input pins of the second LUT. Furthermore we tagged that path with a
MAXDELAY constraint of a few ps to enforce the shortest path between these two
LUTs. After the Xilinx Software finished the place and route process, we
extracted the simulated path length from the logfile. To be sure that exactly
this path will be used in the later design, we used the Xilinx FPGA-Editor to
extract the ROUTE-Constraint of the simulated path.

\begin{figure}
 \includegraphics[width=\textwidth]{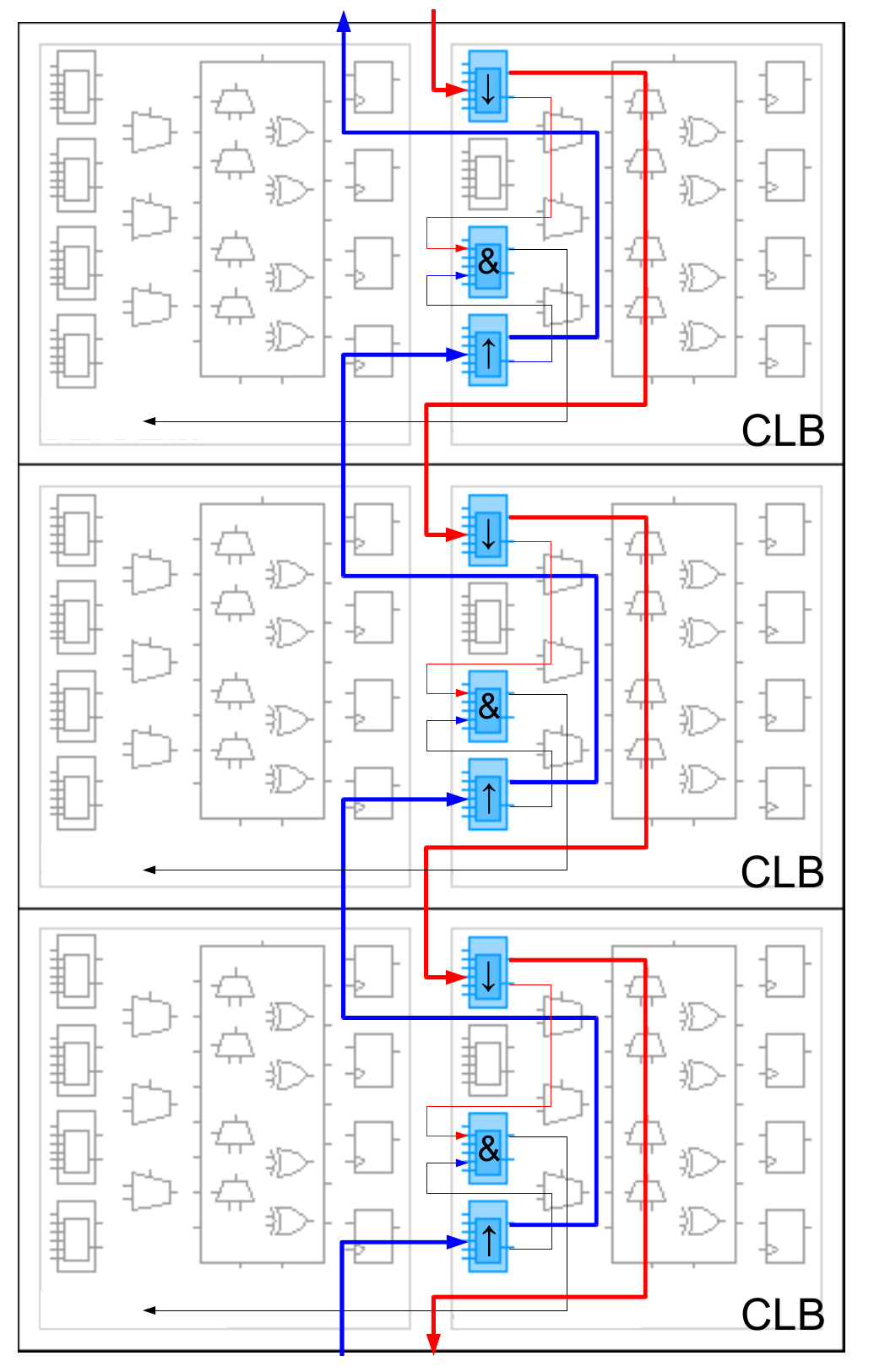}
\caption{Routing of the tapped delay lines and coincidences inside the FPGA
  for one mean-timer. 
The background picture was generated by the ISE software tool ~\cite{ise}. \label{hobclb}}
\end{figure}

This process had been automated to simulate all conceivable combinations and
we found two oppositely running paths between adjacent CLBs with an equal length of
579 ps. However, these paths lengths depend on the absolute position of the used
adjacent CLBs in the CLB grid of the FPGA, but it was possible to identify
enough CLBs with the same simulated timing behavior to construct 64 identical
mean-timers. The resulting design is
illustrated in Fig. \ref{hobclb}.
%The tapped delay lines are therefore based on inter CLB connections
%between adjacent CLBs which are available in a sufficiently large number to reach the
%desired dynamic range.
%The design is illustrated in Fig.~\ref{hobclb}.
%The positions of the LUTs for the tapped delay line inside the CLB had to be chosen 
%in a way that a constant delay between all delay elements was achieved. 
%With the help of simulations all conceivable combinations were tested.
%We found 2$\times$64 TDLs with an identical delay of 579~ps for 64 mean-timers.
The propagating signals in the TDLs are not only sent to the next delay element but also to 
one of the inputs of the ANDs.
These ANDs are realized in one of the 6 remaining LUTs of each CLB (2 are already used
for the tapped delay lines).
The propagation time from the TDL LUTs to the ANDs has to be identical 
only within one mean-timer, differences between different mean-timers can be compensated by delays of each input.

To implement 64 mean-timers in 46 columns with 160 CLBs each, two mean-timers need to be placed 
in one column. We choose to design the tapped delay lines with 53 delays and thus
54 LUTs  which amounts to a total 
range of 53$\times$579~ps $\approx$  30~ns. 
The mean-timers are placed at the top and at the bottom of 32 CLB columns.
This is advantageous for the implementation of the coincidence logic between
the two hodoscopes discussed in Section~\ref{coinc}.

The last step consists of implementing the OR of all 54 AND outputs for each
mean-timer.  As stated before, each LUT has only six inputs, thus the OR has
to be realized as a cascade of multiple ORs.  All signals of the same level of
this cascade should have equal delays. The search for suited combinations has
been performed with an automated script, as for the TDLs. To be able to place all 64
mean-timers side by side, the paths of the cascade are not allowed to leave
the CLB column.  A solution was found using a 6 level cascade with 2 to 1 ORs
where the propagation time variation is at most 34~ps.  These solutions are
different for every mean-timer, leading to different total propagation times
which are again compensated by the input delays of each mean-timer.

\begin{figure}
\includegraphics[width=\textwidth]{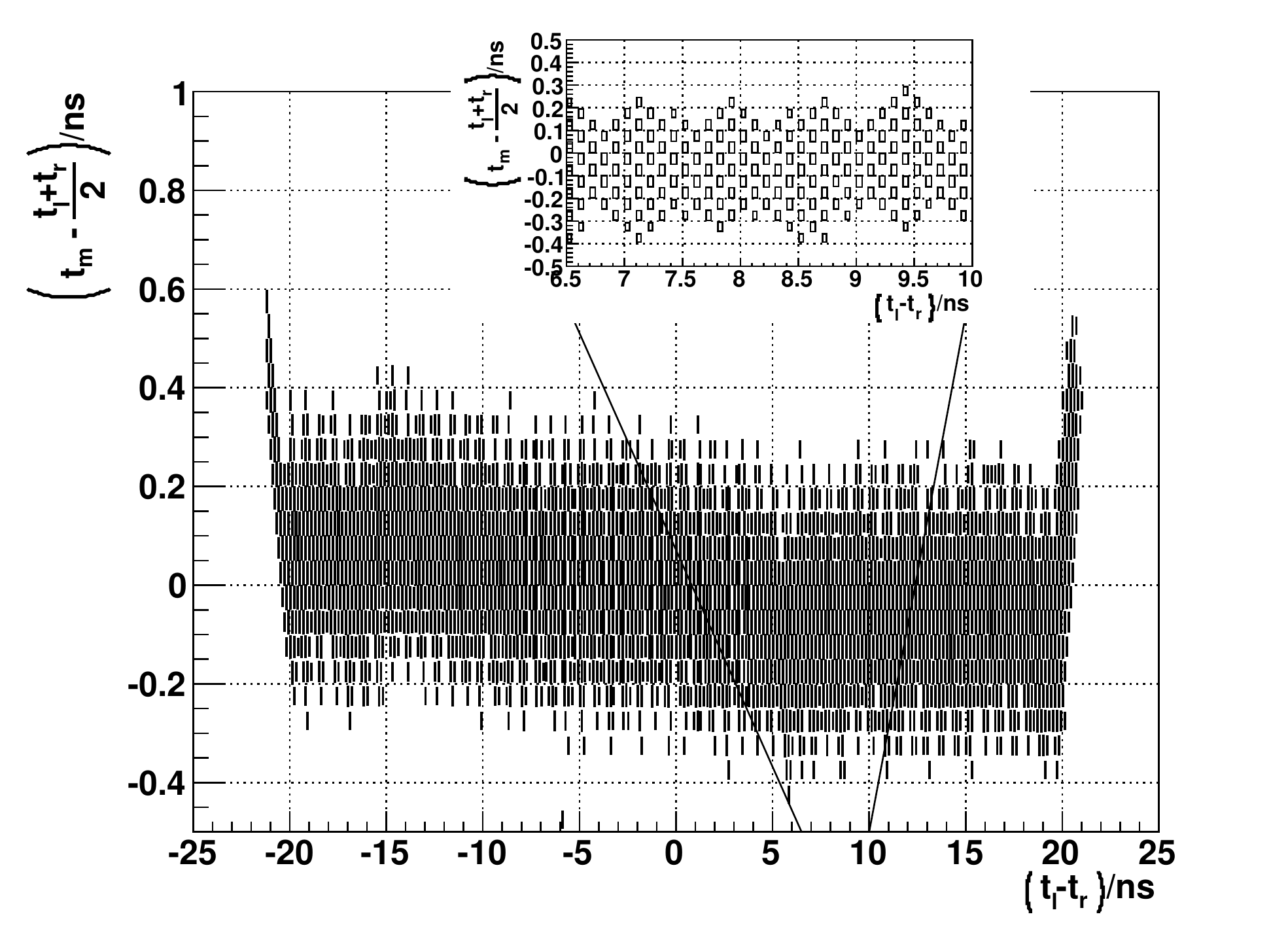}\\
\caption{The difference of the mean-time generated by the FPGA, $t_m$,
and the calculated mean-time $(t_l+t_r)/2$.
 \label{mt_test}}
\end{figure}

Fig.~\ref{mt_test} shows the result of a test measurement
where two input signals
derived from two slightly detuned clocks are sent to the two inputs of one
mean-timer and simultaneously to TDCs for time measurement.
The $x$ axis shows the time difference of the input signals, $t_l-t_r$.
The $y$ axis shows the time difference between the output signal of the 
mean-timer, $t_m$, and the mean-time calculated from $(t_l+t_r)/2$.
The propagation time of the signals through the FPGA
including the coincidence logic 
described in Sec.~\ref{coinc} is approximately 70~ns.
\begin{figure}
\includegraphics[width=\textwidth]{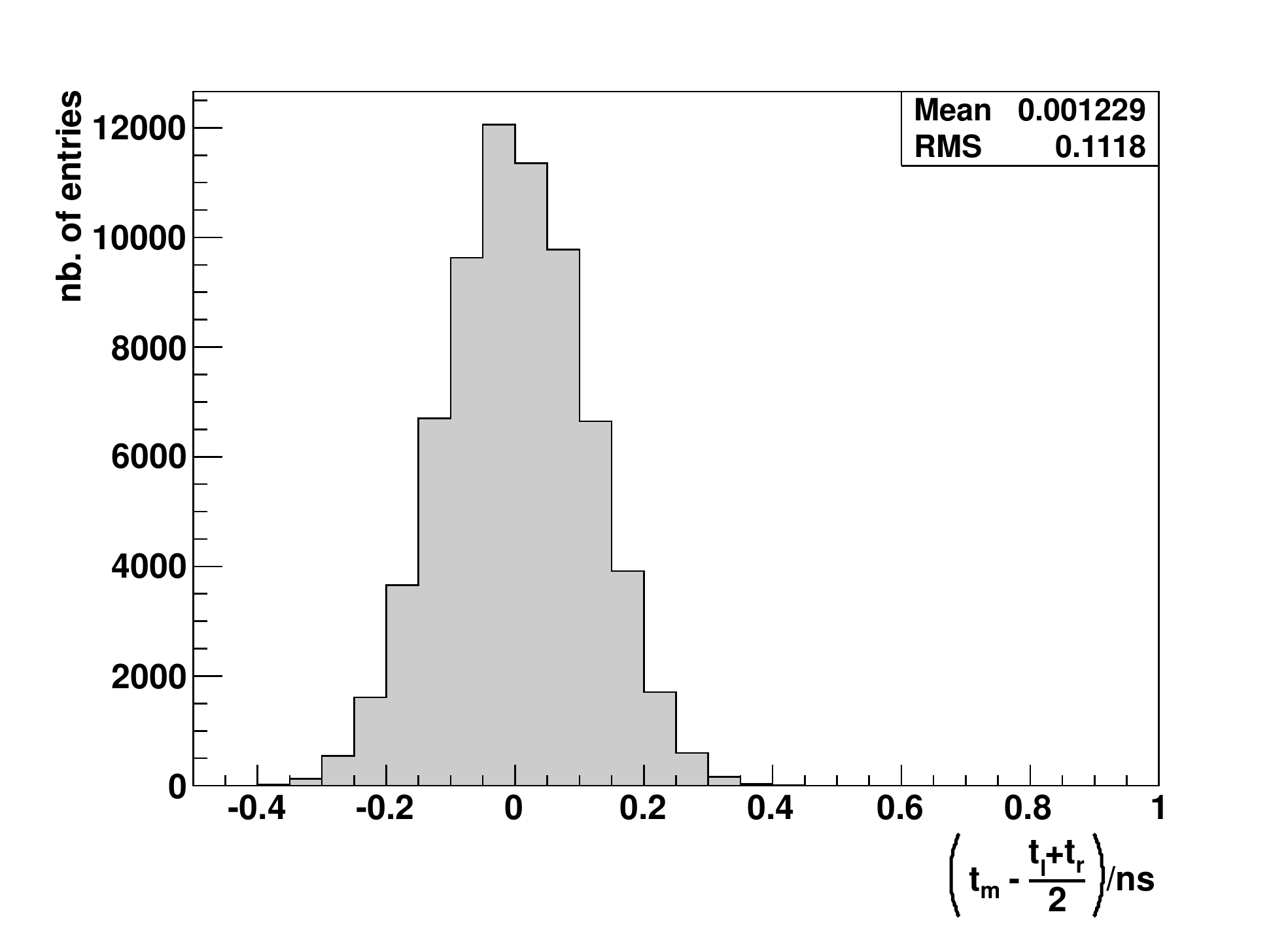}
\caption{Projection of Fig.~\ref{mt_test} on the $y$-axis. \label{mt_test_proj}}
\end{figure}
The output shows the desired behavior in a time range of $\pm$20~ns, i.e.
the mean-time is independent of the time difference $t_l-t_r$.
One also clearly observes a modulation in the inset which shows a magnification on the region 6.5 to 10~ns of the $x$-axis,
due to the tapped delay line.
The observed jitter of the order of 120~ps (RMS) as seen in the projection
on the $y$-axis in Fig.~\ref{mt_test_proj} is acceptable. 
The observed slope of 5~ps per ns originates from a slightly different propagation time in the two TDLs which was not present in the simulation.
At the borders the mean-time signal comes later by up to 200~ps.
The reason for this effect is the following.
If one of the signals already left the tapped
delay line when the second signal just enters, the AND is not
triggered by the leading edge of the signal leaving the tapped delay line, the mean-time signal is
correspondingly delayed.
For the data shown in Fig.~\ref{mt_test} this effect was minimized 
by shortening the input signals inside the FPGA
to a width of 1~ns.

The total dynamic range of $\pm20$~ns is shorter than predicted by simulations ($\pm 30$~ns).
This is due to the simulation software which overestimates the propagation time.
This software is designed for clocked applications for which an upper limit of propagation time must be guaranteed,
however, most signal paths are faster than simulated.
The influence of temperature variation of the FPGA on the propagation time of the signals through
all logic elements inside the FPGA can be neglected.
It was measured to be 30~ps/K in a range from 40 to 80$^\circ$C inside the FPGA chip.

\subsection{Implementation of the coincidence matrix logic}\label{coinc}

\begin{figure}
\includegraphics[width=\textwidth]{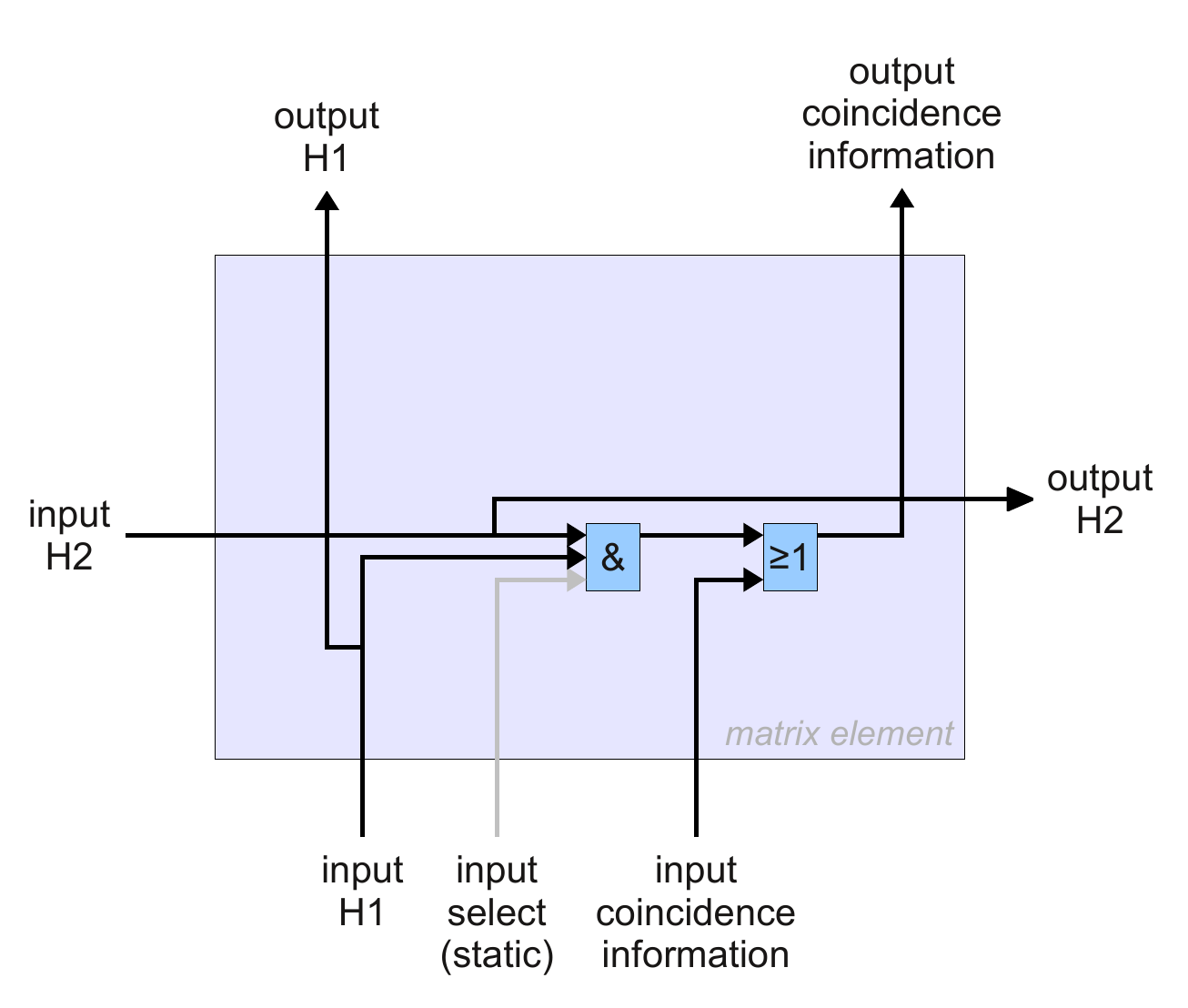}
 \caption{One matrix element. It can be enabled by setting the select input to a logical 1.
The signals H1 and H2 propagate to the corresponding next elements.
The information whether at least one enabled coincidence occurred is
collected in the coincidence information signal. \label{mat_el}}
\end{figure}

The coincidence logic has to be implemented in a way that arbitrary
combinations between 32 inputs from the first hodoscope (H1) with 
32 inputs from the second hodoscopes (H2), i.\,e. in total 1024 ANDs are
selectable in a so called coincidence matrix. 
One of the 1024 matrix elements is shown in Fig.~\ref{mat_el}. The selection is done by using a third input on a AND gate
which is set to a logic 1 if the channel combination is selected. 
The trigger signal is then given by an OR of all 1024 ANDs. 
It is required that the OR signal is generated at a fixed time with respect to
the particle passing the hodoscopes, independently of the  channels
hit. Moreover, for each matrix element the respective signals of the two 
hodoscope counters have to reach the AND at the same time.

\begin{figure}
\includegraphics[width=\textwidth]{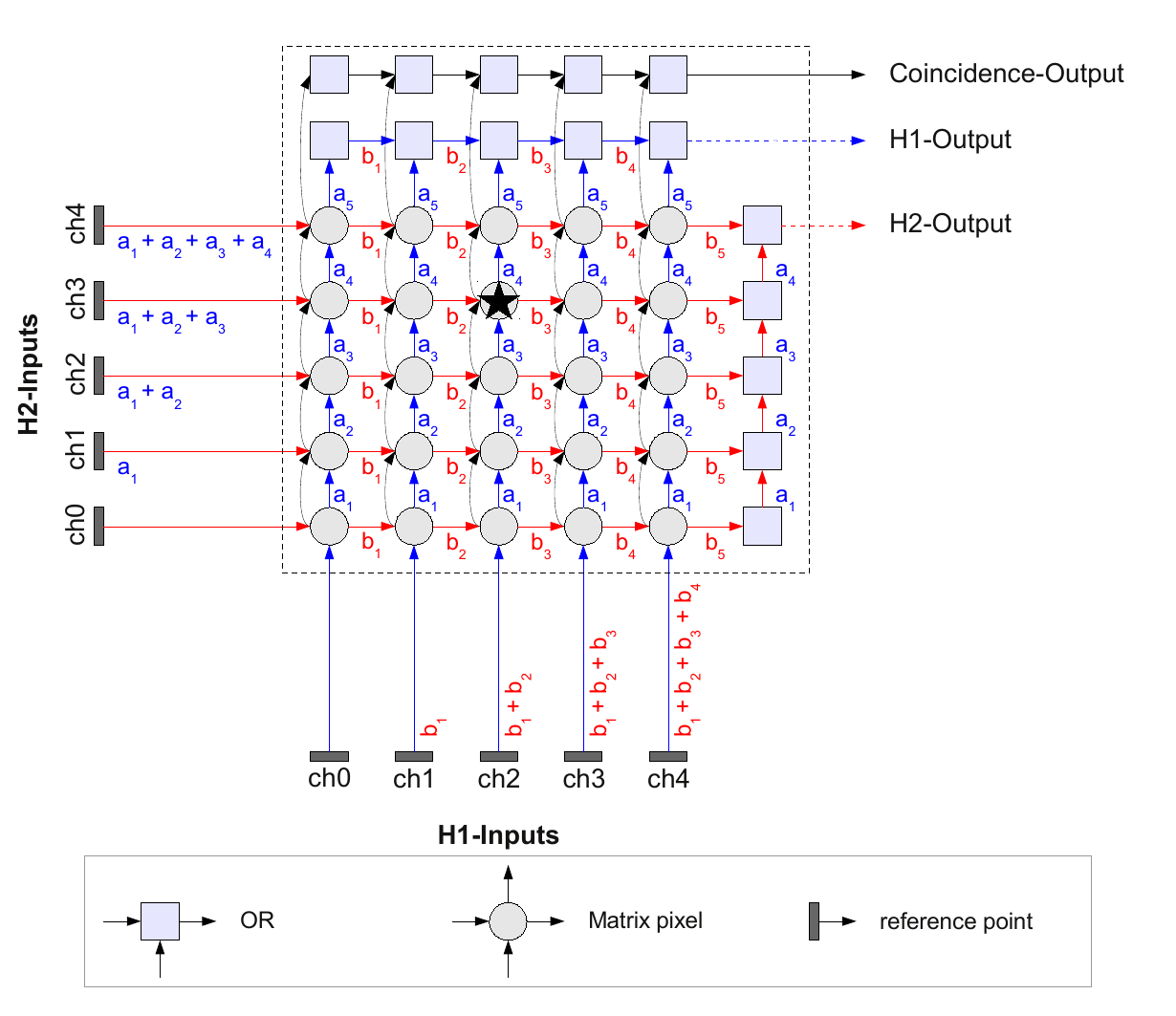}
\caption{
Principle of the coincidence matrix logic between the two hodoscopes. The H1
(H2) output lines represent an OR of all H1 (H2) hodoscope channels, the
coincidence output is the OR of all ANDs and gives the trigger signal
(MATRIX-OUT in Fig.~\ref{system}) after re-timing with the H1-output. 
\label{coinc_matrix}}
\end{figure}

The scheme presented in Fig.~\ref{coinc_matrix} meets both requirements. The $2 \times 32$
signals propagate through the matrix with the delays $a_1, a_2, a_3, \ldots$
and  $b_1, b_2, b_3, \ldots$ which are given by the interconnect paths between
the LUTs on the FPGA. Signal delays between the mean-timer outputs and the
matrix inputs have to be adjusted in order to compensate for the delays inside
the matrix as stated in the figure. 

Consider for example the element marked with $\star$. Both signals are delayed
by $a_1+a_2+a_3+b_1+b_2$ before they arrive at the matrix element. Moreover, it
is easy to see that all signals from input H1 (input H2) reach output H1
(output H2) at the same time. 
The OR of all 1024 AND outputs carries the logic information whether
at least one selected coincidence between H1 and H2 has occurred.
The timing information of this occurrence is derived from the output H1.

\begin{figure}
\includegraphics[width=\textwidth]{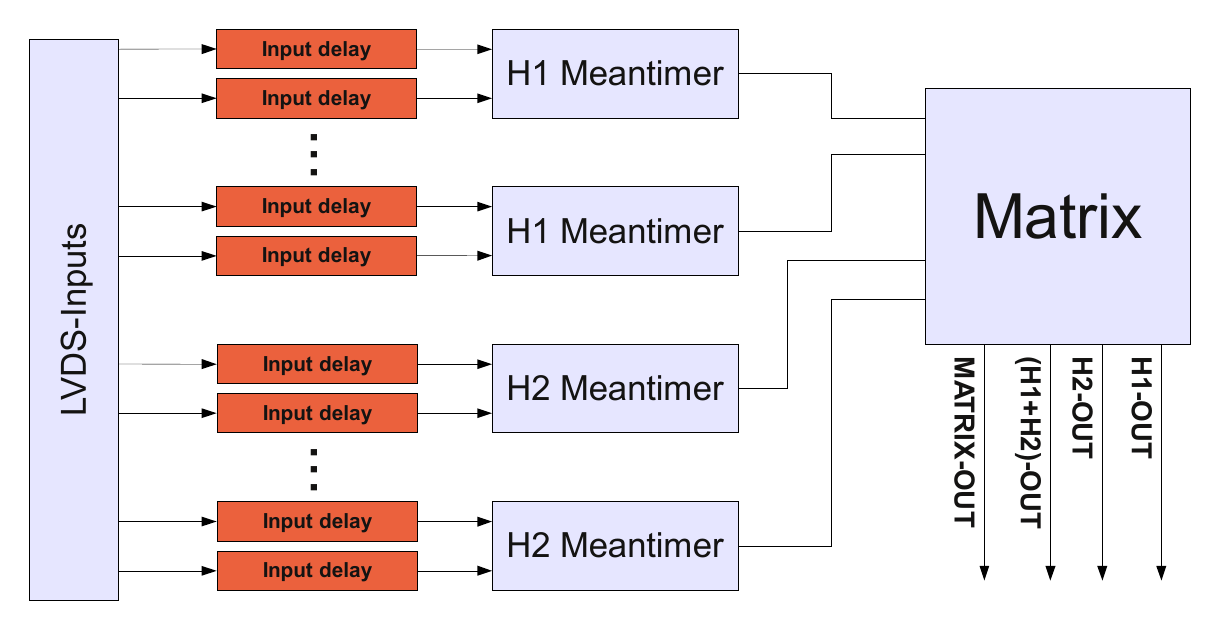}
\caption{The entire system starting from the LVDS inputs to the matrix outputs.\label{system}}
\end{figure}

Fig.~\ref{system} shows schematically the entire system starting from the LVDS
inputs to the outputs. Apart from the output of the matrix needed
for the trigger signal, several auxiliary outputs, like the OR of all channels
in a hodoscope are provided. A web interface allows to set conveniently all delays
and to select the coincidences of the matrix. More details can be found in \cite{john}.

\section{Performance of the entire system}\label{perf}
\begin{figure}
\includegraphics[width=\textwidth]{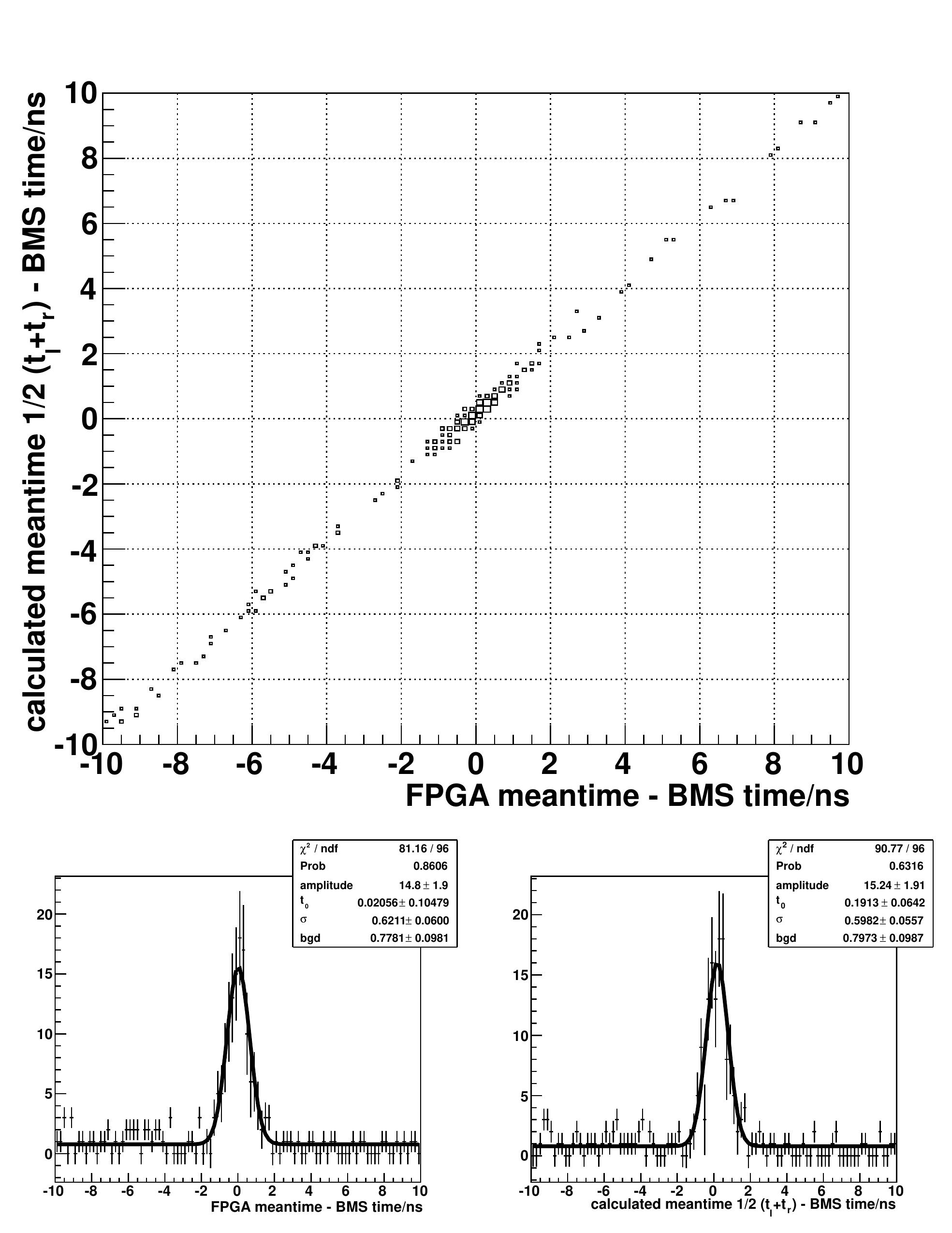}
\caption{The two dimensional histogram shows the mean-time calculated from 
the TDC information of the two PMTs vs. the mean-time
generated by the FPGA.
The two plots on the bottom show the projections on both
axes.\label{mt_plot}}
\end{figure}

The system comprising two GANDALF modules was successfully installed and operated for the 2010 data taking period
at the COMPASS experiment at CERN~\cite{compass} on a newly installed hodoscope pair
enlarging the kinematic range of the trigger system.
Fig.~\ref{mt_plot} demonstrates the performance of the whole system,
starting from the scintillation detectors up to the generation of a trigger signal
in the FPGA.   
The upper plot shows the calculated mean-time from 
the TDC information of the two PMTs of one scintillator element
vs. the mean-time signal generated by the FPGA.
A detector located in the beam line of the experiment
with a good time resolution of 300~ps serves as a reference (``BMS time'').
The strong correlation proves the correct operation of the system.
The two plots on the bottom show the projections on both
axes. Both distributions have approximately the same width of 600~ps,
dominated by the resolution of the scintillation counter.

\begin{figure}
 \includegraphics[width=\textwidth]{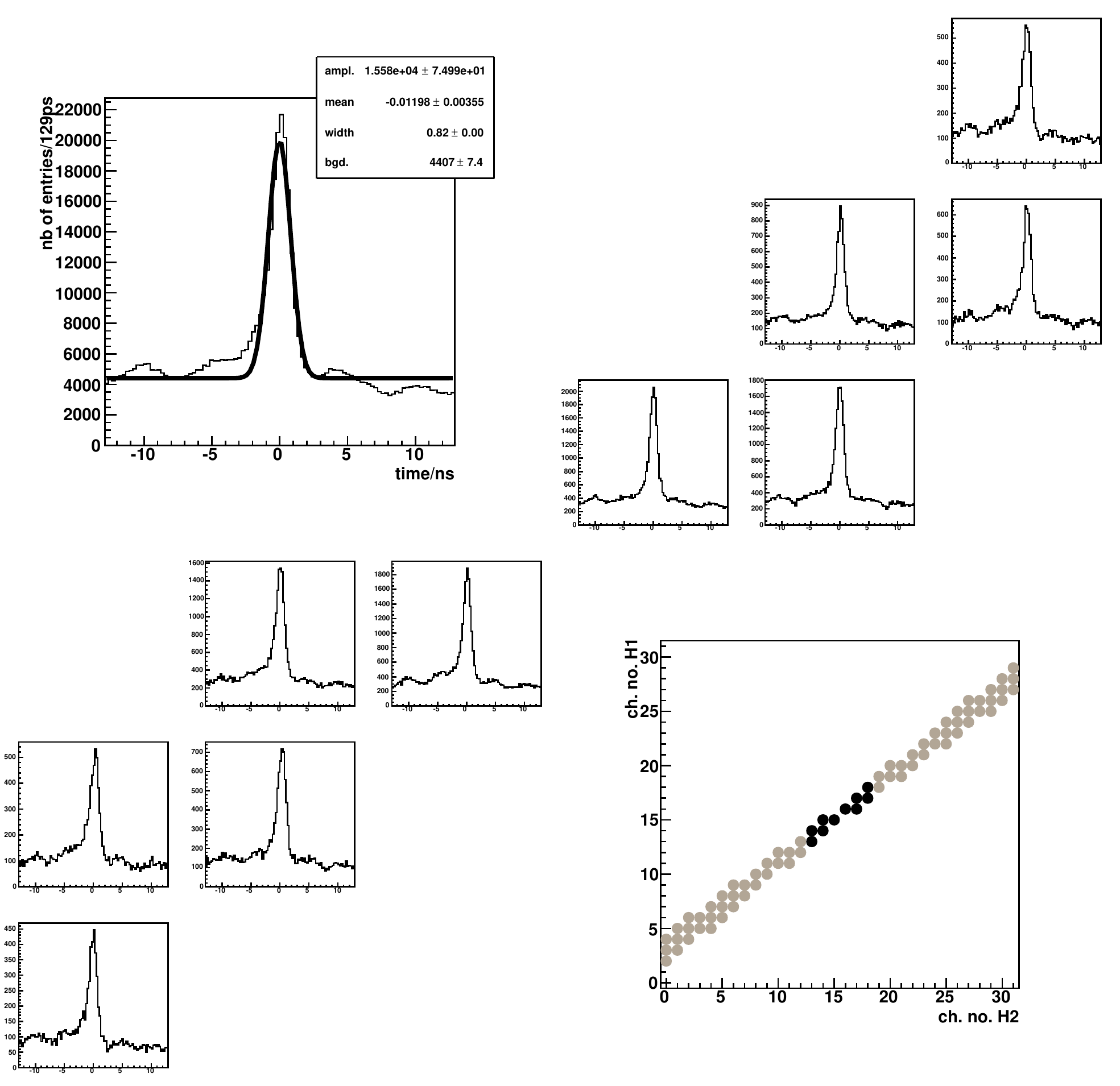}
\caption{Timing of a few matrix elements and timing
of all channels (upper left plot) with respect to a beam line detector.
The lower right plot shows the matrix pattern with the allowed coincidences.
The dark matrix elements correspond to the smaller timing histograms in the figure.
The non-constant background mainly visible in the upper left plot is due
to the time structure of the beam.
\label{timing_LAS1}}
\end{figure}

The upper left plot in Fig.~\ref{timing_LAS1}
shows again the timing of the trigger signal vs. the beam line detector, but
this time all possible coincidences between channels in the two hodoscopes indicated in the matrix 
pattern in the lower right plot contribute to the spectrum. The fact that the width of the Gaussian of 820~ps 
is only slightly worse than the width for the single channel in Fig.~\ref{mt_plot} 
proves that the matrix scheme presented in Section~\ref{coinc} is working and that timing 
adjustments succeeded. Note that the timing adjustments have to be
performed for each FPGA individually to compensate variations of internal delays.
The small spectra in Fig.~\ref{timing_LAS1} show the corresponding time distributions of the central matrix
elements indicated by dark circles in the lower right plot.

\section{Summary}
In summary, a combined mean-time and coincidence logic based on FPGA
technology was developed and implemented at the COMPASS experiment at CERN.
The design was based on a Xilinx Virtex-5 FPGA.
For the first time an unclocked scheme was used for such an implementation.
The mean-timer operates in a range of $\pm$ 20~ns with a resolution of 
approximately 400~ps. On a single FPGA board in total $2 \times 32$ mean-timer
and a 32$\times$32 coincidence logic was implemented.
The spectra shown stem from the 2010 COMPASS data taking period and prove
that the system is running successfully. 

We acknowledge the help of our COMPASS colleagues, especially
Johannes Bernhard, Nicolas du Fresne von Hohenesche and Eva-Maria Kabu{\ss}
who were responsible for the design and installation of the hodoscope system.
The developments described in this report are supported by the German
Bundesministerium f\"ur Bildung und Forschung and the European Community
Research Infrastructure Integrating Activity under the FP7 Study of
Strongly Interacting Matter (HadronPhysics2, Grant Agreement number
227431).

%\clearpage

\appendix
\section{Mean-timing using a tapped delay line}\label{app_tdl}

The basics of the tapped delay line (TDL) technique is outlined in this appendix. 
Denoting by $\Delta_{SL}$ ($\Delta_{SR}$) in Fig.~\ref{meantimer}
the light propagation time from the impact point of the particle in the
scintillator to the left (right) PMT and by $\Delta_{ML}$ ($\Delta_{MR}$)
the corresponding times until both signals meet at one of the ANDs,
the time difference between the occurrence of the event and triggering
of one of the ANDs is 
\[
t= \Delta_{SL} + \Delta_{ML}  =\Delta_{SR} + \Delta_{MR} \, .
\]
Denoting by $\Delta_{S} = \Delta_{SL} + \Delta_{SR}$ the constant time
of light propagation through the entire scintillator
and by  $\Delta_{M} = \Delta_{ML} + \Delta_{MR}$ the constant
total propagation time through the complete tapped delay line,
one finds
\[
 t = \frac{\Delta_S + \Delta_M}{2} = \mbox{const} \, ,
\]  
i.e. the time $t$ when the first AND fires is independent of the differences
in the light propagation time to the left and right end of the scintillator strip.

\end{document}